\documentclass[pra,aps,showpacs,groupedaddress,superscriptaddress,twocolumn,longbibliography,10pt]{revtex4-2}
\usepackage{preamble}
\begin{document}
\title{Extracting transport coefficients from local ground-state currents}

\author{F.~A.~Palm\orcidlink{0000-0001-5774-5546}}
\email{felix.palm@ulb.be}
\affiliation{\ulbaddress}
\affiliation{\solvayaddress}
\author{A. Impertro\orcidlink{0000-0002-0609-4189}}
\affiliation{Fakult\"{a}t f\"{u}r Physik, Ludwig-Maximilians-Universit\"{a}t, 80799 Munich, Germany}
\affiliation{Max-Planck-Institut f\"{u}r Quantenoptik, 85748 Garching, Germany}
\affiliation{Munich Center for Quantum Science and Technology (MCQST), 80799 Munich, Germany}
\affiliation{Institute for Quantum Optics and Quantum Information of the Austrian Academy of Sciences, 6020 Innsbruck, Austria}
\author{M. Aidelsburger}
\affiliation{Fakult\"{a}t f\"{u}r Physik, Ludwig-Maximilians-Universit\"{a}t, 80799 Munich, Germany}
\affiliation{Max-Planck-Institut f\"{u}r Quantenoptik, 85748 Garching, Germany}
\affiliation{Munich Center for Quantum Science and Technology (MCQST), 80799 Munich, Germany}
\author{N.~Goldman\orcidlink{0000-0002-0757-7289}}
\email{nathan.goldman@lkb.ens.fr}
\affiliation{\ulbaddress}
\affiliation{\solvayaddress}
\affiliation{\lkbaddress}
\date{\today}
\maketitle

\section*{Abstract}
\textbf{Transport properties are central to characterizing quantum matter, yet their extraction typically requires external forcing and time-resolved measurements. In this work, we propose a scheme to access transport coefficients directly from measurements of local static ground-state currents -- quantities readily accessible in quantum-engineered platforms. By exploiting the exponential decay of correlations in gapped systems and the finite velocity of correlation spreading, we demonstrate that the local Hall response of correlated insulators can be reconstructed from a small set of quasi-local current observables. We derive explicit relations connecting these static observables to a practical local Chern marker, and introduce a scalable digital protocol for measuring the required generalized currents in quantum simulators. We demonstrate the applicability of our approach through numerical studies of emblematic
Chern-insulator systems, both in the non-interacting and strongly-correlated (fractional) regime. Our method extends naturally to a broad class of correlated systems, even at finite temperature, offering a broadly applicable route to probing transport in engineered quantum matter.}

\section*{Introduction}
\begin{figure}[b!]
    \centering
    \includegraphics{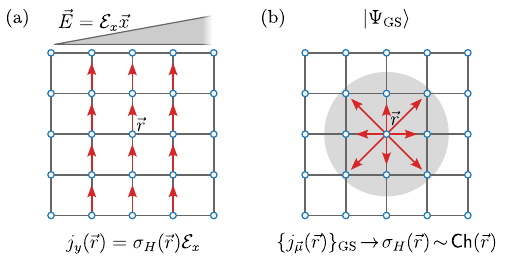}
    \caption{
        (a) Traditional methods to access transport coefficients like the Hall conductivity often use external forcing.
        (b) In contrast, we propose to measure ground-state currents (red arrows) within a finite radius (gray shading) around a reference site (blue dot) to extract the local Hall response (Chern marker).
    }
    \label{fig:Sketch}
\end{figure}

Transport measurements are essential for identifying and characterizing exotic quantum matter, including superconducting states, strange metals, and topological insulators.
While transport is traditionally probed by measuring currents upon connecting reservoirs to a sample, transport coefficients can also be extracted indirectly through other observables and protocols.
Examples of such approaches include the density response to a magnetic perturbation following St\v{r}eda's formula~\cite{Streda1982,Streda1982a}, circular dichroism via Kramers-Kronig relations~\cite{Bennett1965,Tran2017,Repellin2019} and related spectroscopic probes~\cite{Tokuno2011}, as well as center-of-mass drifts upon acting with an external force~\cite{price2012mapping,dauphin2013extracting,Repellin2020}.
These methods have proven to be powerful to probe transport coefficients in various settings, including synthetic quantum matter of cold atoms in continuous systems~\cite{Krinner2017,Amico2022} and optical lattices~\cite{Aidelsburger2015,Asteria2019,Anderson2019,Brown2019,Leonard2023}, as well as photonic systems~\cite{wang2024realization,Chenier2024}.

Theoretically, the conductivity can be expressed as a current-current correlation function via the Kubo formula~\cite{kubo1957statistical}, suggesting that transport could be monitored by evaluating local currents in the ground state.
As further discussed below, this would a priori require monitoring these currents over time, within a region located around a reference site delimited by the correlation length.
While challenging in traditional solid-state platforms, this is particularly appealing to recent quantum-engineered setups, where local currents can be measured with great accuracy~\cite{Impertro2024}.
Local transport properties can thus be extracted, which, for topological insulating states, provides a practical scheme to access local probes of topology, e.g.~local Chern markers~\cite{bianco2011mapping,bianco2013orbital,Seleznev2023,bermond2025local}.
At this stage, however, the measurement of temporal correlations is experimentally challenging; see Ref.~\cite{Knap2013} for a proposal based on many-body Ramsey interferometry.

In this work, we introduce a scheme to extract the local Hall response (Chern marker) of a gapped many-body system, which only requires local  ground-state current measurements. In contrast to more conventional approaches, this scheme entirely builds on static measurements, performed in the absence of any external perturbation.
Importantly, this scheme exploits the exponential decay of correlations~\cite{Hastings2006} in gapped systems, together with the finite velocity of the spread of correlations~\cite{Lieb1972}. Under these conditions, the relevant information can be accessed from a power series of the time-evolved current-current correlations.
As a consequence, the problem reduces to measuring equal-time current correlations within a finite region surrounding the reference site.
Focusing on the transverse (Hall) conductivity, we demonstrate that the associated current–current correlator admits a substantial simplification: it can be expressed in terms of a limited set of further-range current observables. Crucially, we find that static ground-state current measurements alone suffice to accurately determine the Hall conductivity.
This paradigm is summarized schematically in Fig.~\ref{fig:Sketch}.
Inspired by recent experiments in cold-atom quantum simulators~\cite{Impertro2024}, we provide a scalable, digital protocol to access the required further-range currents.
We also present a numerical validation of our scheme, considering a non-interacting Chern insulator in the Harper-Hofstadter model, as well as a strongly-correlated fractional Chern insulator of hard-core bosons.

Since the correlation structures used in our protocol are common to all gapped, topological states, we expect our protocol to apply to a broad class of systems, including systems at finite temperature.
Furthermore, similar protocols can be developed for other correlation functions, giving access to properties not easily accessible in traditional solid-state platforms.
\section*{Results}
\textbf{From the Hall conductivity to local ground-state currents.}\label{section_kubo}
Using Kubo's formula, the local transverse conductivity, evaluated at position $\vec{r}$, reads
\begin{equation}
    \sigma_{xy}(\vec{r}, \omega) = \frac{1}{\omega} \int_0^{\infty}\dd t \int\dd^2r^{\prime}~ \mathrm{e}^{\di\omega t} \braket{\left[\hat{j}_{x}(\vec{r}^{\prime}, t) , \hat{j}_{y}(\vec{r}, 0)\right]},
\end{equation}
where $\omega$ is the frequency of the external perturbing field, and where $\hat{j}_{\alpha}(\vec{r}, t)$ denote the local current operators~\cite{Bruus2004}. In the following, we will assume that the expectation values $\braket{\cdot}$ are evaluated with respect to the system's ground state; we note that the linear-response formalism readily extends to systems described by a mixed-state density matrix~\cite{kubo1957statistical}.

In this work, we focus on the DC transverse (Hall) conductivity,
\begin{equation}
    \sigma_{\rm H}(\vec{r}) = \lim_{\omega\to 0} \Re\sigma_{xy}(\vec{r}, \omega),
    \label{eq:LocalHall_beforeFourier}
\end{equation}
which can be rewritten as
\begin{equation}
    \sigma_{\rm H}(\vec{r}) = \lim_{\omega\to0} \Re \int_0^{\infty}\dd t~ \frac{2\di}{\omega}\de^{\di\omega t} \mathcal{C}(\vec{r}, t),
    \label{eq:FourierTransform_C_of_t}
\end{equation}
where we introduced
\begin{equation}
    \mathcal{C}(\vec{r}, t) \equiv \int\dd^2r^{\prime}~ \Im\braket{\hat{j}_x(\vec{r}^{\prime}, t) \hat{j}_y(\vec{r}, 0)},
    \label{eq:CorrelatorC_of_t}
\end{equation}
and used the relation
\begin{equation}
    \braket{\left[\hat{A}, \hat{B}\right]} = 2\di \Im \braket{\hat{A} \hat{B}}
\end{equation}
for Hermitian operators.

\textbf{The single-frequency ansatz.}
The unequal-time correlation function contained in Eq.~\eqref{eq:CorrelatorC_of_t} cannot be accessed in  state-of-the-art quantum simulators. For gapped states, we can circumvent this limitation by approximating the correlator  $\mathcal{C}(\vec{r}, t)$ with an exponentially damped oscillation,
\begin{equation}
    \mathcal{C}(\vec{r}, t) \approx \mathcal{C}(\vec{r}, 0) \de^{-\Gamma t} \cos(\omega_0 t).
    \label{eq:Approximate_C_of_t}
\end{equation}
This approximation is applicable under the assumption that there is a single characteristic gap $\omega_0$, from the ground state to the relevant low-energy excitations. The rate $\Gamma$ then characterizes the broadening of this transition due to band dispersion and finite-size effects. This single-frequency ansatz is expected to hold in the regime of relatively flat Bloch bands, as we explicitly demonstrate below and in the Methods.
%

Importantly, the ansatz in Eq.~\eqref{eq:Approximate_C_of_t} factorizes the time-dependence of the correlator $\mathcal{C}(\vec{r}, t)$. It thus simplifies the task to measuring \textit{equal-time} current correlations $\mathcal{C}(\vec{r}, 0)$ -- readily accessible across various platforms -- alongside determining the characteristic gap $\omega_0$, which can be effectively probed via spectroscopy. It is worth mentioning that equal-time current correlations can be exactly connected to density correlations in  Landau-level states~\cite{SpasicMlacak2025}, hence suggesting further practical simplifications in this special case.

Using the approximate form of the correlator in Eq.~\eqref{eq:Approximate_C_of_t}, one can analytically evaluate the Fourier transform in Eq.~\eqref{eq:FourierTransform_C_of_t}, which yields 
\begin{equation}
    \sigma_{\rm H}(\vec{r}) = 2 \mathcal{C}(\vec{r}, 0) \frac{\omega_0^2 - \Gamma^2}{\left(\Gamma^2 + \omega_0^2\right)^2} \equiv {\sf Ch} (\vec{r})/2\pi.
    \label{eq:HallFromDampedOscillations}
\end{equation}
This expression offers an explicit recipe to obtain the local Hall response from the three parameters entering the single-frequency ansatz in Eq.~\eqref{eq:Approximate_C_of_t}: $\mathcal{C}(\vec{r}, 0)$, $\omega_0$ and $\Gamma$. Since the Hall response is directly related to the Chern number~\cite{Thouless1982,Niu1985}, we point out that Eq.~\eqref{eq:HallFromDampedOscillations} provides a simple and practical expression for a local Chern marker ${\sf Ch} (\vec{r})$.

\begin{figure}[t!]
    \centering
    \includegraphics{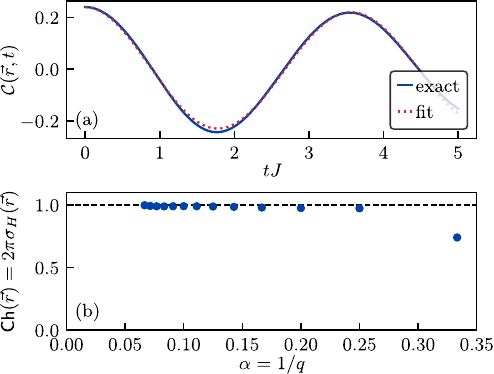}
    \caption{
        (a) In a Chern insulator state, the current-current correlator $\mathcal{C}(\vec{r}, t)$ defined in  Eq.~\eqref{eq:CorrelatorC_of_t}, relative to a reference site $\vec{r}$, exhibits damped oscillations, which are well captured by the ansatz in Eq.~\eqref{eq:Approximate_C_of_t}.
        (b) Upon fitting the damped oscillations, one can use the expression in Eq.~\eqref{eq:HallFromDampedOscillations} to obtain a local Chern marker. The result is in  agreement with the Chern number ${\sf Ch}=1$ of the populated band. Here, the calculations were performed by filling the lowest band of the Harper-Hofstadter model with non-interacting fermions, at flux $\alpha=\nicefrac{1}{q}$ per plaquette.
    }
    \label{fig:NumericalBenchmark}
\end{figure}

To confirm the validity of this approximation, we performed a numerical study of a paradigmatic Chern insulator state, which is obtained by filling the lowest band of the Harper-Hofstadter model with non-interacting fermions; see also the next paragraph
for  details.
We find that the correlator $\mathcal{C}(\vec{r},t)$ that was obtained numerically is in good agreement with the ansatz in Eq.~\eqref{eq:HallFromDampedOscillations}; see Fig.~\ref{fig:NumericalBenchmark}(a).  Here, we extracted the parameters $\Gamma$ and $\omega_0$ by fitting the numerical data. As discussed in detail below, these parameters match the bandwidth $\Gamma\!\approx\!W$ and cyclotron gap $\omega_0\!\approx\!\Omega_c$ of the corresponding Hofstadter bands, respectively.

Furthermore, we confirm that the local Chern marker ${\sf Ch} (\vec{r}) = 2\pi \sigma_{\rm H}(\vec{r})$ [Eq.~\eqref{eq:HallFromDampedOscillations}], extracted from the fitted curve, agrees well with the expected Chern number \mbox{${\sf Ch}=1$} of the lowest Hofstadter band in the limit of sufficiently small $\alpha$; see Fig.~\ref{fig:NumericalBenchmark}(b).
Deviations at large $\alpha$ are attributed to the stronger dispersion of the Hofstadter bands in this regime, which broadens the transition.

We explicitly confirm the local nature of the Chern marker ${\sf Ch} (\vec{r})$ in the Methods,
where we present the results for a junction displaying two (local and distinct) Chern insulating phases.

 \textbf{Unequal-time correlators and static ground-state observables.}
 As already remarked, it is challenging to measure the unequal-time correlator $\braket{\hat{j}_{x}(\vec{r}, t) \hat{j}_{y}(\vec{r}^{\prime}, 0)}$ defined in Eq.~\eqref{eq:CorrelatorC_of_t}, both in quantum-simulator experiments and in numerical studies of strongly-correlated systems. And while the ansatz in Eq.~\eqref{eq:Approximate_C_of_t} offers a possible simplification to the problem, it would be appealing to identify a general, practical and self-contained method to systematically determine such unequal-time correlators without any a priori knowledge of the system's characteristics (e.g.~the spectral gap $\omega_0$).
To do so, we proceed by rewriting the formal expression for this correlator in a more appealing form, using the Baker-Campbell-Hausdorff identity, 
\begin{align}
    \mathcal{C}(\vec{r}, t)
    &= \int\dd^2r^{\prime}~ \Im\braket{\de^{\di\hcH t} \hat{j}_x(\vec{r}^{\prime}) \de^{-\di\hcH t} \hat{j}_y(\vec{r})} \\
        &= \int\dd^2r^{\prime}~ \Im\braket{  \hat{j}_x(\vec{r}^{\prime}) \left( \sum_{m=0}^{\infty} \frac{(-\di t)^m}{m!} \left[\hcH, \hat{j}_y(\vec{r})\right]_m\right) },\notag
\end{align}
where $\left[\cdot, \cdot\right]_m$ denotes an $m$-fold commutator.
We can further rewrite this expression as
\begin{equation}
    \mathcal{C}(\vec{r}, t) = \sum_{m=0}^{\infty} \frac{t^m}{m!} c_m,
    \label{eq:C_of_t_Expansion}
\end{equation}
where we introduced the coefficients
\begin{equation}
    c_m \equiv\Im\left[ (-\di)^m \int\dd^2r^{\prime}~ \braket{\hat{j}_x(\vec{r}^{\prime}) \left[\hcH, \hat{j}_y(\vec{r})\right]_m} \right] .
    \label{eq:coefficients_cm}
\end{equation}
We emphasize that the coefficients $c_m$ are entirely expressed as static (ground-state) observables.

For sufficiently short-ranged interactions, correlations spread with a finite velocity~\cite{Lieb1972}, which in combination with the finite correlation length of a gapped system allows us to extract the characteristic features of $\mathcal{C}(\vec{r}, t)$ already from its early-time behavior.
In this scenario, it is sufficient to consider a finite set of coefficients $c_m$ in Eq.~\eqref{eq:C_of_t_Expansion},
\begin{equation}
    \mathcal{C}(\vec{r}, t) \simeq c_0 + c_1 t + \frac{c_2}{2} t^2 + \mathcal{O}(t^3).
    \label{eq:C_of_t_Expansion:2ndOrder}
\end{equation}
Indeed, in combination with the ansatz in Eq.~\eqref{eq:Approximate_C_of_t}, one finds that only three coefficients suffice, with the correspondence given by
\begin{equation}
    \mathcal{C}(\vec{r}, 0) = c_0, \quad \Gamma = -\frac{c_1}{c_0}, \quad \omega_0 = \sqrt{\left(\frac{c_1}{c_0}\right)^2 - \frac{c_2}{c_0}}.
    \label{eq:ConnectionCm_DampedOsci}
\end{equation}
In this way, static ground-state properties [Eq.~\eqref{eq:coefficients_cm}] give direct access to the Hall response in Eq.~\eqref{eq:HallFromDampedOscillations} without the need for time-evolution nor transport measurements.

Importantly, we remark that $c_1$ scales with the damping $\Gamma$, hence it is small whenever band-dispersion effects are limited.
In fact, in our numerical investigations detailed below, we find that $c_1$ is indeed negligible, as well as higher odd corrections ($c_{3, 5, \hdots}$).
In contrast, we find that including even corrections ($c_{4, 6, \hdots}$) improves the determination of the correlator $\mathcal{C}(\vec{r}, t)$; see the Methods.
However, including those higher-order corrections leads to a significant overhead due to the numerous currents that need to be evaluated; see below.

Furthermore, we remind that the coefficient $\omega_0$, which requires the measurement of $c_2$, corresponds to the characteristic gap in the flat-band regime. Thus, this coefficient could be evaluated independently through conventional spectroscopy. Altogether, this indicates that the measurement of the transport coefficient \mbox{${\sf Ch} (\vec{r}) = 2\pi \sigma_{\rm H}(\vec{r})$} can be limited to the evaluation of the single coefficient $c_0$ in the flat-band regime (i.e.~the low-flux limit in the Harper-Hofstadter model).

Since the explicit form of the coefficients $c_m$ depends on the details of the Hamiltonian, we will now investigate concrete examples in the next paragraph, based on the Harper-Hofstadter model.
However, we expect our approach to apply to a broad class of gapped, local Hamiltonians.

\textbf{Extracting the $c_m$ coefficients in the Harper-Hofstadter model.}
%
We consider the Harper-Hofstadter model, defined on a square lattice, with flux $\alpha$ per plaquette. In the Landau gauge, the Hamiltonian reads
\begin{equation}
    \hcH\!=\!-J \sum_{x,y} \left(\hcd_{x+1,y} \hc_{x,y} + \de^{\di 2\pi \alpha x} \hcd_{x,y+1} \hc_{x,y} + \mathrm{h.c.}\right),
\end{equation}
where $\vec{r}=(x,y)$ denote the lattice sites, $J$ is the hopping amplitude, and $\hat{c}^{(\dagger)}_{x,y}$ are annihilation (creation) operators for fermions or bosons, depending on the context below.
This well-known model exhibits topological bands~\cite{Thouless1982}, with Chern number ${\sf Ch}=1$ in the lowest band for $\alpha\!=\!1/q$, with $q$ an integer. The continuum limit of the model is reached upon setting $\alpha \rightarrow 0$, where the Hofstadter bands reduce to (flat) Landau levels.

%
Here, our aim is to obtain a concrete and practical expression for the coefficients $c_m$ defined in Eq.~\eqref{eq:coefficients_cm} in terms of local ground-state currents, in view of reconstructing the local Hall response through simple static measurements [Eqs.~\eqref{eq:HallFromDampedOscillations} and \eqref{eq:ConnectionCm_DampedOsci}]. To do so, we first identify the conserved currents in our lattice model, using the continuity equation for the local number operator $\hn_{\vec{r}}\!=\! \hcd_{\vec{r}} \hc_{\vec{r}}$,
    \begin{align}
        \frac{\dd}{\dd t} \braket{\hn_{\vec{r}}} &= \di \braket{\left[\hcH, \hn_{\vec{r}}\right]}\\
        &= - \left( \braket{\hat{j}_{\hat{x}}(\vec{r}; 0)} - \braket{\hat{j}_{\hat{x}}(\vec{r}-\hat{x}; 0)} \right.\notag\\
        &\phantom{=-\left(\right.}\left.+ \braket{\hat{j}_{\hat{y}}(\vec{r}; 2\pi\alpha x)} - \braket{\hat{j}_{\hat{y}}(\vec{r}-\hat{y}; 2\pi\alpha x)}\right).\notag
    \end{align}
Here, $\hat{x}$ and $\hat{y}$ denote the unit vectors along the respective direction and we have introduced the generalized current operators
\begin{equation}
    \hat{j}_{\vec{\mu}}(\vec{r}; \theta) = iJ \mathrm{e}^{i\theta} \hat{c}^{\dagger}_{\vec{r}+\vec{\mu}} \hat{c}^{\vphantom\dagger}_{\vec{r}} + \mathrm{H.c.} ,
\end{equation}
which describe the conserved currents on the links of the lattice, but also the longer-range currents that will appear to be relevant in our protocol below.

Using these current operators, we can now calculate the commutators with the Hamiltonian entering Eq.~\eqref{eq:coefficients_cm}: 
\begin{widetext}
\begin{equation}
    \begin{aligned}
        &\left[\hcH, \hat{j}_{\vec{\mu}}(\vec{r}; \theta)\right]\\
        &=iJ \left(
            \hat{j}_{\vec{\mu}+\hat{x}}(\vec{r}; \theta+\frac{\pi}{2}) + \hat{j}_{\vec{\mu}+\hat{x}}(\vec{r}-\hat{x}; \theta+\frac{3\pi}{2}) + \hat{j}_{\vec{\mu}+\hat{y}}(\vec{r}; \theta+2\pi\alpha(x+\mu_x)+\frac{\pi}{2}) + \hat{j}_{\vec{\mu}+\hat{y}}(\vec{r}-\hat{y}; \theta+2\pi\alpha x+\frac{3\pi}{2})
            \right.\\
        &\phantom{=iJ \left(\right.}\left.
            + \hat{j}_{\vec{\mu}-\hat{x}}(\vec{r}; \theta+\frac{\pi}{2}) + \hat{j}_{\vec{\mu}-\hat{x}}(\vec{r}+\hat{x}; \theta+\frac{3\pi}{2}) + \hat{j}_{\vec{\mu}-\hat{y}}(\vec{r}; \theta-2\pi\alpha(x+\mu_x)+\frac{\pi}{2}) + \hat{j}_{\vec{\mu}-\hat{y}}(\vec{r}+\hat{y}; \theta-2\pi\alpha x+\frac{3\pi}{2})
        \right).
    \end{aligned}
    \label{eq:mFoldCommutator}
\end{equation}
\end{widetext}
Considering the $m$-fold commutator in Eq.~\eqref{eq:coefficients_cm}, we find that the coefficients $c_m$ only involve simple expectation values of the form
\begin{equation}
    \Im\braket{\hat{j}_{\hat{x}}(\vec{r}^{\prime}; \theta=0) \hat{j}_{\vec{\mu}}(\vec{R}; \Theta)},
    \label{eq:ExpectationValues_CurrentCorrs}
\end{equation}
where the sets of relevant $\left(\vec{\mu}, \vec{R}, \Theta\right)$ are generated by repeatedly applying the identity in Eq.~\eqref{eq:mFoldCommutator} starting from the reference site $\left(\hat{y}, \vec{r}, 2\pi\alpha x\right)$.

As a key step in our approach, we then notice that the quantities in Eq.~\eqref{eq:ExpectationValues_CurrentCorrs} can be further simplified as a sum of local current expectation values,
%
\begin{equation}
    \begin{aligned}
        \Im&\braket{\hat{j}_{\hat{x}}(\vec{r}^{\prime}) \hat{j}_{\vec{\mu}}(\vec{R};\Theta)}\\
        &= \frac{J}{2} \left( \braket{\hat{j}_{\vec{\mu}+\hat{x}}(\vec{R}; \Theta)} \delta(\vec{r}^{\prime}-(\vec{R}+\vec{\mu}))\right.\\
        &\phantom{=\frac{J}{2}~~}\left.~- \braket{\hat{j}_{\vec{\mu}+\hat{x}}(\vec{R}-\hat{x}; \Theta)} \delta(\vec{r}^{\prime}-(\vec{R}-\hat{x}))\right.\\
        &\phantom{=\frac{J}{2}~~}\left.~- \braket{\hat{j}_{\vec{\mu}-\hat{x}}(\vec{R}; \Theta)} \delta(\vec{r}^{\prime}-(\vec{R}-\hat{x}+\vec{\mu})) \right.\\
        &\phantom{=\frac{J}{2}~~}\left.~+ \braket{\hat{j}_{\vec{\mu}-\hat{x}}(\vec{R}+\hat{x}; \Theta)} \delta(\vec{r}^{\prime}-\vec{R}) \right).
    \end{aligned}
    \label{Eq:ConnectionRule}
\end{equation}
Importantly, this indicates that all the coefficients $c_m$ in Eq.~\eqref{eq:coefficients_cm}, which are required to extract the local Hall response [Eqs.~\eqref{eq:HallFromDampedOscillations} and \eqref{eq:ConnectionCm_DampedOsci}], can be obtained from generalized current measurements.
We emphasize that this is yet another substantial simplification compared to the initial task to measure the explicit time-evolution of a current-current correlator.
In fact, below we will provide a scalable, digital protocol to access these quasi-local currents.

Finally, the coefficients $c_m$ in Eq.~\eqref{eq:coefficients_cm} can be determined by applying the simplification scheme described above [Eqs.~\eqref{eq:mFoldCommutator}-\eqref{Eq:ConnectionRule}] to the $m$-fold commutator in Eq.~\eqref{eq:coefficients_cm}. The final expression for these coefficients thus formally reads
\begin{equation}
    c_m = J^m \sum_{\left(\vec{\mu}_m, \vec{R}_m, \Theta_m\right)} \Im\braket{\hat{j}_{\hat{x}}(\vec{r}^{\prime}) \hat{j}_{\vec{\mu}_m}(\vec{R}_m, \Theta_m)},
    \label{eq:currentCurrentCorrs_simplified}
\end{equation}
where each term corresponds to a set of finite-range current expectation values [Eq.~\eqref{Eq:ConnectionRule}].

\begin{figure}[t!]
    \centering
    \includegraphics{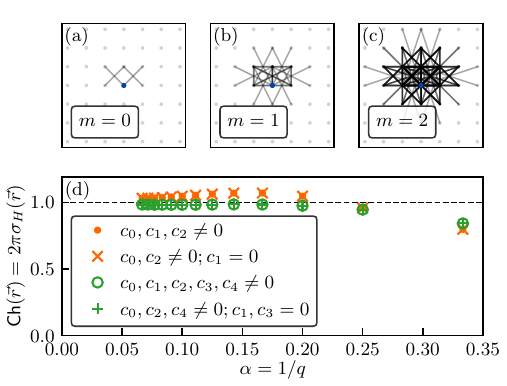}
    \caption{
        (a-c) Currents $\hat{j}_{\vec{\mu}}(\vec{R}, \Theta)$ to be evaluated at (a) zeroth, (b) first, and (c) second order in $t$ for a Harper-Hofstadter model on a square lattice of $13\times 13$ sites with the reference site (blue) chosen in the center of the system.
        Darker arrows indicate that more than one generalized current with different phases $\Theta$ have to be measured.
        The necessary currents are found by first applying the iterative construction in Eq.~\eqref{eq:mFoldCommutator}, then extracting the imaginary part of the correlator using Eq.~\eqref{Eq:ConnectionRule}, and integrating over all lattice sites $\vec{R}$.
        (d) Local Chern marker ${\sf Ch}(\vec{r}) = 2\pi\sigma_{\rm H}(\vec{r})$ [Eq.~\eqref{eq:HallFromDampedOscillations}] as obtained from the coefficients $c_m$ in Eq.~\eqref{eq:coefficients_cm} using the experimental protocol.
        %
        Here, the coefficients $c_m$ are connected to the parameters of Eq.~\eqref{eq:Approximate_C_of_t} via Eq.~\eqref{eq:ConnectionCm_DampedOsci}.
        The values for the Chern marker obtained in this manner are in good agreement among each other as well as with the expected ${\sf Ch}=1$ for the lowest Hofstadter band.
        For additional details see the Methods section.
    }
    \label{fig:ExpansionApproach}
\end{figure}

We note that the number of summands in Eq.~\eqref{eq:currentCurrentCorrs_simplified}
is given by $8^m$ [Eq.~\eqref{eq:mFoldCommutator}], with each summand contributing $4$ generalized currents [Eq.~\eqref{Eq:ConnectionRule}].
Therefore, at $m$-th order $4 \times 8^m$ currents need to be measured, in total resulting in \mbox{$4\times (8^0 + 8^1 + 8^2) = 292$} current measurements to access the local Hall conductivity from the leading-order coefficients $c_{0,1,2}$. In Fig.~\ref{fig:ExpansionApproach}(a-c) we show the currents that have to be evaluated for a Harper-Hofstadter model on a $13\times13$ square lattice with the reference site chosen at the center of the system.

We close this paragraph by reminding that the measurement of the transport coefficient ${\sf Ch} (\vec{r}) = 2\pi \sigma_{\rm H}(\vec{r})$ can be limited to the evaluation of the single coefficient $c_0$, in the flat-band regime; see above.
This measurement would only require the determination of the four local diagonal currents displayed in Fig.~\ref{fig:ExpansionApproach}(a), which is experimentally appealing.

\textbf{Numerical study I: Non-interacting Chern insulator.}
 We demonstrate the applicability of our scheme through a numerical study of an emblematic Chern-insulator system, obtained by considering non-interacting fermions in the Harper-Hofstadter model.
In our simulations, we consider an open square of size $L \times L = 13\times13$ with varying flux $\alpha=\nicefrac{1}{q}$, $q=3, \hdots, 15$, per plaquette, so that the lowest band has Chern number ${\sf Ch}=1$.
We construct a Chern insulator by occupying the lowest $N=\nicefrac{(L_x \times L_y)}{q}$ states with fermions and use this state to test our experimental protocol.
In particular, we implement the recursive generation of the currents visualized in Fig.~\ref{fig:ExpansionApproach}(a-c) and evaluate their expectation values to obtain $c_{0, 1, 2}$.
From these, we can extract the parameters of the damped oscillation [Fig.~\ref{fig:NumericalBenchmark}(a)] through Eq.~\eqref{eq:ConnectionCm_DampedOsci} and insert them into the expression for the local Chern marker ${\sf Ch} (\vec{r})$ in Eq.~\eqref{eq:HallFromDampedOscillations}.

We find that the results obtained in this manner are in good agreement with the expected Chern number ${\sf Ch}=1$ of the lowest Hofstadter band, see Fig.~\ref{fig:ExpansionApproach}(d).
We emphasize again that, in this approach, the problem reduced to the evaluation of $292$ generalized ground-state currents, instead of performing explicit time-evolution.
While including the higher order contribution $c_4$ further improves the results, we find that neglecting the odd orders ($c_1$ and $c_3$) does not affect the results; see also the Methods.

\begin{figure}
    \centering
    \includegraphics{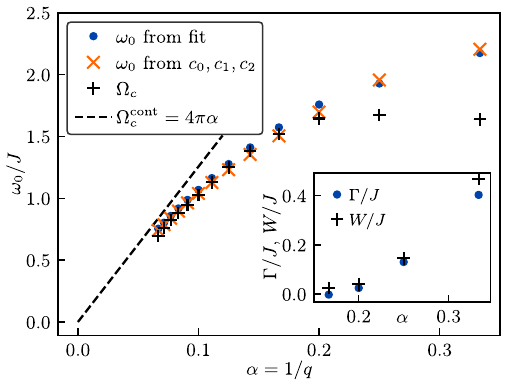}
    \caption{
        Characteristic frequency $\omega_0$ obtained from a fit of the correlator $\mathcal{C}(\vec{r}, t)$ using Eq.~\eqref{eq:Approximate_C_of_t} (blue dots), and compared to that extracted from the coefficients $c_m$ in Eq.~\eqref{eq:ConnectionCm_DampedOsci} (orange crosses). The cyclotron frequency $\Omega_c$ obtained from the single-particle spectrum is also indicated.
        The bandwidth $W$ of the lowest Hofstadter band is plotted in the inset.
        It is shown to agree with the decay rate of the correlations $\Gamma \approx W$. 
        The cyclotron frequency $\Omega_c$ approaches the continuum value $\Omega_c^{\rm cont}=4\pi\alpha$ in the low-flux (continuum) limit.
    }
    \label{fig:Gap}
\end{figure}

%
Before turning to a second application of our method, we comment on the applicability of the single-frequency ansatz in Eq.~\eqref{eq:Approximate_C_of_t}.
We compare the characteristic oscillation frequency $\omega_0$, extracted from a fit of the correlator $\mathcal{C}(\vec{r}, t)$ using Eq.~\eqref{eq:Approximate_C_of_t}, to the cyclotron gap $\Omega_c$.
The latter quantity corresponds to the bandgap to the first excited band in the Harper-Hofstadter model, and reaches the value $\Omega_c^{\rm cont}=4\pi\alpha$ in the low-flux (continuum) limit.
As shown in Fig.~\ref{fig:Gap}, the extracted oscillation frequency $\omega_0$ agrees well with the cyclotron frequency $\Omega_c$ over a broad flux window, $\alpha \lesssim 0.2$, hence validating the single-frequency ansatz in Eq.~\eqref{eq:Approximate_C_of_t} in this range.
 We also show that the parameter $\omega_0$ extracted from the coefficients $c_m$ [see Eq.~\eqref{eq:ConnectionCm_DampedOsci}] agrees with the single-frequency ansatz [Eq.~\eqref{eq:Approximate_C_of_t}] throughout the entire flux range.
The identification of the cyclotron bandgap as the relevant frequency indicates that the parameter $\omega_0$ can be determined independently through spectroscopy.

In contrast, for $\alpha\gtrsim0.2$ we find significant deviations between the cyclotron gap $\Omega_c$ and the fitted $\omega_0$.
This agrees with the substantial band dispersion in this regime, resulting in a finite bandwidth $W/J$.
We note that this broadening of the lowest band results in an exponential decay of the correlations with a rate $\Gamma \approx W$; see the inset in Fig.~\ref{fig:Gap}.

\textbf{Numerical study II: Laughlin state of hard-core bosons.}
We complete our numerical validation by exploring the applicability of our method to a bosonic Laughlin state~\cite{Laughlin1983}:~a strongly-correlated topological state.
Specifically, we consider the Harper-Hofstadter model discussed above, and add repulsive Hubbard interactions of strength $U>0$,
\begin{equation}
    \hcH_{\rm int} = \frac{U}{2} \sum_{\vec{r}} \hn_{\vec{r}} (\hn_{\vec{r}} - 1).
\end{equation}
For simplicity, we consider the hard-core bosonic limit ($U/J\to \infty$), where the exact expressions for the commutators in Eq.~\eqref{eq:mFoldCommutator} remain unchanged.
At magnetic filling factor $\nu=\nicefrac{1}{2}$, it is well-established that this model hosts a lattice analog of the corresponding Laughlin state~\cite{Soerensen2005}.

For large systems, while ground-state searches and evaluations of (quasi-)local currents are numerically feasible, long-time evolution of the strongly-correlated many-body system becomes increasingly challenging.
Therefore, we apply the protocol outlined above to extract the local Chern marker from static ground-state currents.
To this end, we employ matrix product states (MPS) and the density-matrix renormalization group (DMRG) to variationally find the ground state~\cite{White1992,Schollwoeck2011}, using the MPSKit toolkit~\cite{MPSKit}.

\begin{figure}[b!]
    \centering
    \includegraphics{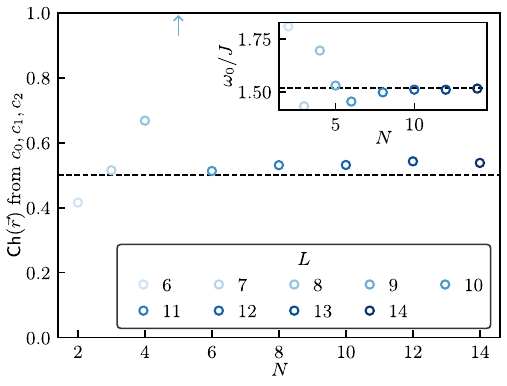}
    \caption{
        Local many-body Chern marker $\mathsf{Ch}(\vec{r})$ [Eq.~\eqref{eq:HallFromDampedOscillations}] for a reference site in the center of an $L\times L$-system of $N$ hard-core bosons at $\alpha=\nicefrac{1}{6}$ in the Laughlin regime.
        The marker is obtained via Eq.~\eqref{eq:ConnectionCm_DampedOsci} by evaluating the coefficients $c_0, c_1,$ and $c_2$ in Eq.~\eqref{eq:coefficients_cm} from the static ground-state.
        We find that for sufficiently large $N\geq 6$, the extracted Chern marker is in good agreement with the expected $\sf{Ch}=\nicefrac{1}{2}$ of the Laughlin state (dashed line).
        The inset shows the extracted frequency $\omega_0$, which is in good agreement with the cyclotron frequency of the non-interacting model for large $N$ (dashed line).
    }
    \label{fig:Laughlin}
\end{figure}

For concreteness, we fix the flux $\alpha=\nicefrac{1}{6}$ per plaquette and vary the number of particles $N$ and the system size $L$ to realize a $\nicefrac{1}{2}$-Laughlin state.
We measure the currents indicated in Fig.~\ref{fig:ExpansionApproach}(a-c) and use them to obtain $c_{0,1,2}$ and -- through Eq.~\eqref{eq:ConnectionCm_DampedOsci} -- the local many-body Chern marker $\mathsf{Ch}(\vec{r})$ for a reference site in the center of the system.
We find that for sufficiently large systems ($N\geq 6$, $L\geq 10$) the extracted results converge and are in good agreement with the expected \mbox{$\mathsf{Ch}\!=\!\nicefrac{1}{2}$} of the Laughlin state; see Fig.~\ref{fig:Laughlin}.
We attribute the remaining deviations from the expected value even for large systems to band-dispersion effects due to the relatively large flux $\alpha=\nicefrac{1}{6}$; see also Fig.~\ref{fig:ExpansionApproach}(d).
We remark that the characteristic frequency $\omega_0$ matches well the cyclotron gap of the system; see the inset of Fig.~\ref{fig:Laughlin}.
We attribute this agreement to residual Landau-level selection rules, which are expected to become exact in the low-flux limit of the Harper-Hofstadter model; see also Ref.~\cite{unal2024quantized}.

Our findings indicate that, despite the simplifying assumptions and strong correlations of the Laughlin state, our approach is capable of efficiently extracting a local topological marker with a satisfactory level of accuracy. This is due to the finite correlation length present in the Laughlin state, originating from the many-body bulk gap. This key property enables the establishment of a local Chern marker in strongly-correlated topological states, as similarly demonstrated by the local St\v{r}eda response~\cite{Repellin2020}.

\textbf{Protocol for measuring further-range currents.}
The experimental protocol presented above requires the measurement of generalized, further-range currents connecting lattice sites $\vec{R}$ and $\vec{R}+\vec{\mu}$.
Here, we present a scalable, digital protocol to achieve such measurements using only nearest-neighbor tunnelings $\hat{T}_{ij}$ and local energy offsets $\hat{M}_i$ on the lattice,
\begin{equation}
    \hat{T}_{ij} = -J \left(\ket{i}\bra{j} + \mathrm{H.c.}\right), \qquad \hat{M}_{i} = \Delta \ket{i}\bra{i},
\end{equation}
as well as local density measurements on the sites involved.

We aim at evaluating the current $\hat{j}_{\vec{r}_1, \vec{r}_N}$, from a reference (initial) site $\vec{r}_1 = \vec{R}$ to a final site $\vec{r}_N = \vec{R}+\vec{\mu}$. Having identified a possible path of nearest-neighbor links, we propose to implement a pulse sequence of the form
\begin{widetext}
\begin{equation}
    \hat{U}(t_2) = \exp\left[-\di \frac{\pi}{4J} T_{\vec{r}_{N-1},\vec{r}_N}\right] \exp\left[-\di \frac{\pi}{2J} T_{\vec{r}_{N-2},\vec{r}_{N-1}}\right] ... \exp\left[-\di \frac{\pi}{2J} T_{\vec{r}_1, \vec{r}_2}\right] \exp[-\di t_2\hat{M}_{\vec{r}_N}].
\end{equation}
\end{widetext}
Indeed, by appropriately choosing the duration $t_2$ of the local energy offset $\hat{M}_{\vec{r}_N}$, we find that the population imbalance at the initial and final lattice sites -- upon completing the pulse sequence -- directly relates to the target current through the simple relation
\begin{equation}
\begin{aligned}
    \hat{n}_{\vec{r}_N}(t_{\rm final})-\hat{n}_{\vec{r}_1}(t_{\rm final}) &= \hat{U}^{\dagger}(t_2) \left[\hat{n}_{\vec{r}_N}(0)-\hat{n}_{\vec{r}_1}(0)\right] \hat{U}(t_2) \\
    &= \frac{1}{2J} \hat{j}_{\vec{r}_1, \vec{r}_N}(0) + \sum_{\vec{r}_i} \gamma_i \hat{n}_{\vec{r}_i}(0),
\end{aligned}
\end{equation}
where $\hat{j}_{\vec{r}_1, \vec{r}_N}(0) = \hat{j}_{\vec{\mu}}(\vec{R}; \Theta=0)$ in the earlier notation.
The sum involves the sites along the path, and the $\gamma_i$ are coefficients that have to be determined for the specific path and time $t_2$ chosen; see below.
We note that the length of the pulse sequence is not determined by the Euclidean distance on the lattice, but rather by the number $\ell$ of sites encountered in a path connecting the sites.

\begin{figure}[b!]
    \centering
    \includegraphics{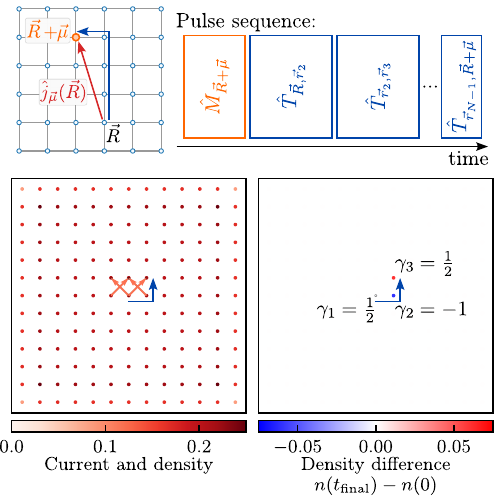}
    \caption{
        Upper panel: Sketch of the pulse sequence needed to evaluate further-range currents.
        Lower left: Ground state density (dots) and $m=0$ currents (red arrows) of the Chern insulator in the lowest band of a Harper-Hofstadter model at \mbox{$\alpha=\nicefrac{1}{5}$}.
        The blue arrow indicates a possible path to measure a current between next-nearest neighbors.
        Lower right: Density difference after the pulse sequence and coefficients $\gamma_i$ with which the respective points along the path are weighted.
    }
    \label{fig:PulseSequence}
\end{figure}

For concreteness, we consider the case of currents between sites connected by paths of two, three, or four sites ($\ell=2,~3,~4$): 
\begin{itemize}
    \item \textbf{$\ell=2$}: This case reduces to the existing protocol of Impertro~\textit{et~al.}~\cite{Impertro2024}, and does not require us to apply the $\hat{M}_2$-gate, i.e. $t_2=0$, nor additional density measurements of the initial state, \mbox{$\gamma_1 = \gamma_2 = 0$}.
    \item \textbf{$\ell=3$}: Here, we need a pulse of duration $t_2 = \frac{\pi}{2\Delta}$ to obtain the required phase (up to a global sign).
    The initial ground-state densities contribute with \mbox{$\gamma_1\!=\!\gamma_3\!=\! \frac{1}{2}$ and $\gamma_2\!=\!-1$} along the path; see Fig.~\ref{fig:PulseSequence}.
    \item \textbf{$\ell=4$}: Here, we again do not need to apply any pulse, $t_2=0$, (up to a global sign),
    however the initial ground-state densities along the path contribute with \mbox{$\gamma_1 = \gamma_4 = \frac{1}{2}$, $\gamma_2 = -1$, $\gamma_3 = 0$}.
\end{itemize}
This pattern repeats as one considers longer paths.
To imprint an additional phase in the generalized current operator, it is sufficient to change the duration of the $\hat{M}_{\vec{r}_N}$-pulse such that it incorporates the correct relative phase.
The pulse sequence as well as an example for a next-nearest neighbor current is exemplified in Fig.~\ref{fig:PulseSequence}.

\section*{Discussion}
In this work, we showed that measurements of ground-state currents of gapped many-body systems are sufficient to extract the local Hall marker.
In particular, using the quasi-local nature of correlations in gapped quantum many-body systems together with the finite velocity of the spread of correlations, we showed that the relevant information is already encoded in the short-time behavior of the correlation function.
Furthermore, we discuss how static ground-state currents encode the same information, thereby providing a concrete experimental protocol to extract transport coefficients from ground-state currents alone.
We confirmed our general arguments by numerical simulations of the paradigmatic Chern insulator state of non-interacting fermions in a completely filled Hofstadter band, as well as the strongly-correlated bosonic Laughlin state in a half-filled Hofstadter band.

From an experimental perspective, while measuring hundreds of further-range currents might be challenging using existing platforms~\cite{Impertro2024}, we identified a realistic scheme based on the evaluation of a single ($c_0$) coefficient, which only requires four local diagonal current measurements.

Owing to its broad applicability, our approach can also be utilized for mixed states, allowing for the investigation of transport coefficients at finite temperatures.
Furthermore, our protocol extends to other (potentially multi-point) correlators, involving (local) densities and currents, thereby giving access to response functions hardly accessible in traditional solid-state experiments.

\section*{Methods}
\textbf{Oscillation frequency of $\mathcal{C}(\vec{r}, t)$.}\label{app:Gap}
To justify our statement that the oscillations of the current-current correlator
\begin{equation}
    \mathcal{C}(\vec{r}, t) = \int\dd^2r^{\prime}~ \Im\braket{\hat{j}_x(\vec{r}^{\prime}, t) \hat{j}_y(\vec{r}, 0)}
\end{equation}
are at a frequency $\omega_0$ related to the gap, we expand the expression in terms of Bloch states $\ket{u_{a}(\vec{k})}$, were $a$ denotes the band.
We assume that the lowest band ($a=0$) is completely filled, whereas the other bands are completely empty.
We find
\begin{widetext}
\begin{equation}
\begin{aligned}
    \mathcal{C}(\vec{r}, t)
    &= \int\dd^2 r^{\prime}~ \sum_{b\geq 1} \int\frac{\dd^2 k}{(2\pi)^2} \Im\left(\braket{u_0(\vec{k}) | \hat{j}_{x}(\vec{r}^{\prime}, t) | u_b(\vec{k})} \braket{u_b(\vec{k}) | \hat{j}_{y}(\vec{r}, 0) | u_0(\vec{k})} \right)\\
    &= \int\dd^2 r^{\prime}~ \sum_{b\geq 1} \int\frac{\dd^2 k}{(2\pi)^2} \Im\left(\de^{\di \left(\epsilon_a(\vec{k}) - \epsilon_b(\vec{k})\right) t} \braket{u_0(\vec{k}) | \hat{j}_{x}(\vec{r}^{\prime}) | u_b(\vec{k})} \braket{u_b(\vec{k}) | \hat{j}_{y}(\vec{r}) | u_0(\vec{k})} \right),
\end{aligned}
\end{equation}
where we introduces the Bloch energies $\epsilon_{a}(\vec{k})$.
To simplify our notation, we introduce the symbol
\begin{equation}
    \Phi_{0,b}(\vec{k}; \vec{r}) = \int\dd^{2} r^{\prime}~ \braket{u_0(\vec{k}) | \hat{j}_{x}(\vec{r}^{\prime}) | u_b(\vec{k})} \braket{u_b(\vec{k}) | \hat{j}_{y}(\vec{r}) | u_0(\vec{k})},
\end{equation}
\end{widetext}
so that we can rewrite the correlator as
\begin{equation}
    \mathcal{C}(\vec{r}, t) = \sum_{b\geq 1} \int\frac{\dd^2 k}{(2\pi)^2} \Im \left( \de^{\di \left(\epsilon_a(\vec{k}) - \epsilon_b(\vec{k})\right) t} \Phi_{0b}(\vec{k}; \vec{r}) \right).
\end{equation}
Assuming that the bands are sufficiently flat so that $\epsilon_{b}(\vec{k}) - \epsilon_0(\vec{k}) = \Delta_{b0}(\vec{k}) \approx \Delta_{b0}$, we see that the correlator has oscillations at frequencies corresponding to the band gap $\Delta_{b0}$ from the lowest band to the $b$-th excited band, i.e.
\begin{equation}
    \mathcal{C}(\vec{r}, t) = \sum_{b\geq 1}\Im\left( \de^{\di \Delta_{b0} t} \left(\int_{FBZ} \Phi_{0b}(\vec{k})\right) \right).\label{multi-gap}
\end{equation}

In principle, the expression in Eq.~\eqref{multi-gap} encompasses multiple energy gaps, resulting in oscillations at various frequencies. Nevertheless, the dominant contribution is expected to arise from transitions between the ground band and the first excited band, thereby producing a single characteristic frequency, $\omega_0$. This behavior is exemplified in the Landau-level limit, where selection rules restrict transitions from the lowest Landau level ($n\!=\!0$) to the next ($n\!=\!1$). Additional selection rules may emerge from specific symmetries, as seen in the Harper-Hofstadter model, which exhibits chiral symmetry~\cite{Repellin2019}.

The validity of this argument can be confirmed by comparing the characteristic oscillation frequency $\omega_0$, extracted from a fit of the correlator $\mathcal{C}(\vec{r}, t)$ using Eq.~\eqref{eq:Approximate_C_of_t}, to the cyclotron gap $\Omega_c$, i.e. the bandgap to the first excited band in the Harper-Hofstadter model, and it reaches the value $\Omega_c^{\rm cont}=4\pi\alpha$ in the low-flux (continuum) limit; see Fig.~\ref{fig:Gap} in the Results section.

\begin{figure}[t!]
    \centering
    \includegraphics{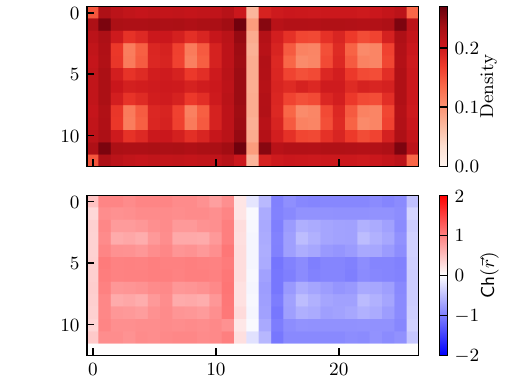}
    \caption{
        Density and local Chern marker as obtained from the fit in Eq.~\eqref{eq:Approximate_C_of_t} for a Harper-Hofstadter model with opposite flux $\pm\alpha=\nicefrac{1}{5}$ per plaquette in the left and right half separated by a single column of a local potential.
        As expected, we find that the local Chern marker ${\sf Ch}(\vec{r})$ changes sign as the reference site $\vec{r}$ is moved from one region to the other.
    }
    \label{fig:Locality}
\end{figure}

\textbf{Locality of the Chern marker.}\label{app:Locality}
To confirm the local nature of the Chern marker, we consider a variant of the Harper-Hofstadter model on an elongated rectangle of size $27 \times 13$ with flux \mbox{$\alpha=\pm\nicefrac{1}{5}$} in the left (right) half, respectively, separated by a single column of local potential $J \sum_{y=1}^{13} \hn_{(14,y)}$ in the middle.
By numerically evaluating $\mathcal{C}(\vec{r},t)$ [Eq.~\eqref{eq:CorrelatorC_of_t}] and extracting the local Hall response $\sigma_{\rm H}(\vec{r})$ from a fit with Eq.~\eqref{eq:Approximate_C_of_t}, we find that the local Chern marker ${\sf Ch}(\vec{r}) = 2\pi\sigma_{\rm H}(\vec{r})$ takes values close to $+1$ ($-1$) in the left (right) half of the system according to the sign of the magnetic flux, see Fig.~\ref{fig:Locality}, thus directly confirming its local nature.

\textbf{Including higher coefficients $c_m$.}\label{app:HigherCms}
In order to improve the accuracy of our protocol, it is natural to include higher coefficients in the expansion of the correlator $\mathcal{C}(\vec{r}, t)$ in Eq.~\eqref{eq:C_of_t_Expansion},
\begin{equation}
    \mathcal{C}(\vec{r}, t) = \sum_{m=0}^{\infty} \frac{t^m}{m!} c_m.
\end{equation}
When including higher coefficients $c_m$, it is also natural to adapt the ansatz for the correlator in Eq.~\eqref{eq:Approximate_C_of_t} to include additional frequencies from, for example, higher bands.
A simple generalization is
\begin{equation}
    \mathcal{C}(\vec{r}, t) \approx \mathcal{C}(\vec{r}, 0) \mathrm{e}^{-\Gamma t} \left(a \cos(\omega_0 t) + (1-a) \cos(2\omega_0 t)\right),
    \label{eq:GeneralizedAnsatz}
\end{equation}
which we found to improve the fit of the signal obtained from exact time evolution.

Upon extracting the coefficients $c_m$ obtained in this manner, we find that
\begin{equation}
    \frac{c_1}{c_0} = -\Gamma, \quad \text{and} \quad c_{m} \propto c_1 \text{ for } m \text{ odd},
\end{equation}
so that under the assumption of limited band dispersion, i.e. small $\Gamma$, the contributions from odd terms can be neglected.
We confirmed this numerically for the original ansatz in Eq.~\eqref{eq:Approximate_C_of_t} (orange data in Fig.~\ref{fig:ExpansionApproach}(d)) and for the extended ansatz in Eq.~\eqref{eq:GeneralizedAnsatz}.
Our findings indicate that incorporating the extra coefficient $c_4$ significantly enhances the extracted Chern marker (as shown by the green data in Fig.~\ref{fig:ExpansionApproach}(d)). However, this comes with a considerable additional overhead due to the necessity of evaluating 16,384 additional currents.

\bibliography{bibliography.bib}

\begin{acknowledgments}
    \ 
    The authors thank Immanuel Bloch for fruitful discussions.
    This research was financially supported by the ERC Grant LATIS, the FRS-FNRS (Belgium), the EOS project CHEQS, and the Fondation ULB.
    A.I. and M.A. acknowledge support from the Deutsche Forschungsgemeinschaft (DFG, German Research Foundation) under Germany’s Excellence Strategy – EXC-2111 – 390814868, the German Federal Ministry of Education and Research via the funding program quantum technologies – from basic research to market (contract number 13N15895 FermiQP) and the Horizon Europe programme HORIZON-CL4-2022-QUANTUM-02-SGA via the project 101113690 (PASQuanS2.1).
\end{acknowledgments}

\end{document}